# Continuous monitoring of membrane protein micro-domain association during cell signaling

Heng Huang[1] and Arnd Pralle[1]

Department of Physics, University at Buffalo, the State University of New York, Buffalo, NY 14260, Correspondence to: Arnd Pralle[1] e-mail: apralle@buffalo.edu

**Central to understanding membrane bound cell signaling is to quantify how the membrane ultra-structure consisting of transient spatial domains modulates signaling and how the signaling influences this ultra-structure. Yet, measuring the association of membrane proteins with domains in living, intact cells poses considerable challenges. Here, we describe a non-destructive method to quantify protein–lipid domain and protein cytoskeleton interactions in single, intact cells enabling continuous monitoring of the protein domains interaction over time during signaling.**

The cell membrane is thought to contain structural and functional subdomains[1-3]: (i) the membrane-associated cytoskeleton forms "fences" which coral diffusing proteins[2], and (ii) sterols and sphingolipids form "lipid rafts"[3-5]. This ultra-structure modulates the diffusion and aggregation of proteins, thereby influencing membrane bound cell signaling[6-10]. However, these structures are too small and possibly too dynamic for visualization by optical microscopy. Thus far, methods used to characterize lipid domains in cells either perturb the domains (detergent extraction[5], choleratoxin or antibody labeling[4,11]), or are technically challenging (single particle tracking, thermal noise imaging and homo-FRET measurements[12-14]). Also, none of these methods are able to continuously measure a protein-lipid raft association or observe changes caused by external stimuli or cell signaling in real time.

Here, we demonstrate a nondestructive method capable of continuously monitoring membrane protein interactions with the membrane ultra-structure. This real-time capability permits us to study changes in protein- structural domain associations as a consequence of signaling, ligand binding, or membrane perturbations such as anesthetics or temperature. We present examples demonstrating changes in protein-lipid domain interaction caused by ion channel gating, antibody cross-linking and cholesterol extraction.

Our method further develops an approach of Wawrezinieck et al.[18] who showed it is possible to characterize membrane protein-lipid domain interactions using a series of diffusion measurements across areas of increasing size. Wawrezinieck employed confocal fluorescence correlation spectroscopy (FCS) to measure diffusion which requires multiple measurements with a series of excitation volumes recorded sequentially, thereby prohibiting real-time investigation of dynamic events.

In contrast, our method utilizes spatially resolved camera based FCS[19] to record the diffusion time $\tau_{1/2}$ across multiple sizes of areas *simultaneously*. The signal obtained in each camera pixel is binned into super-pixels covering a range of sizes. This method, which we call binned-imaging FCS (bimFCS), provides a fast and convenient method to continuously characterize membrane protein diffusion allowing, for the first time, quantification of changes in the protein-lipid raft association in real-time during. Also in contrast to confocal FCS, bimFCS does not require any experimental calibration because the observation area of each pixel is fixed.

Figure 1 describes the principle (**Fig.1 a, b**) of bimFCS and how it effectively characterizes the diffusion of a membrane protein as (**i**) free Brownian diffusion in a homogenous environment, (**ii**) diffusion slowed by transient trapping in sub-microscopic domains, or (**iii**) diffusion hindered by interaction with a grid of diffusion barriers, i.e. 'membrane fences'. A region of the cell membrane is illuminated using objective based total internal reflection (TIR) microscopy. The fluorescence of a green fluorescent protein (GFP) fused to the membrane protein of interest is recorded using an electron multiplying charge-coupled device (EMCCD). To measure the transit time $\tau_{1/2}$ of the membrane protein through different area sizes simultaneously, the camera pixels are binned into super-pixels from bin 1 to 6, which spans from a diffraction limited spot to a 0.96µm x 0.96µm area (For instrument details see supplementary information (**Fig. S1**).

The bimFCS measurement allows us to distinguish between three types of membrane protein diffusion by examining the relationship between $\tau_{1/2}$ and $\omega^2$, which quantifies the size of the observation area: (**i**) For a protein diffusing freely in a homogenous membrane the relationship is expected to be linear and to intercept through the origin. (**ii**) If the $\tau_{1/2}$ versus $\omega^2$ relationship is linear but intercepts the time axis at a positive value the diffusion is hindered by transient, sub-microscopic traps smaller than the smallest measurement area, i.e. lipid rafts. (**iii**) For a membrane protein which interacts with 'membrane fences' the $\tau_{1/2}$ versus $\omega^2$ relationship displays a steepening of the slope above a typical area size $\omega^2$. The area $\omega^2$ at which the slope increases indicates the average free area between the filaments. This behavior is quantified by linearly extrapolating the steeper slope of the long range diffusion, i.e. the larger super-pixels, to the time axis yielding a negative time axis intercept. In all cases, the inverse slope of the curve is a measure of the effective diffusion constant $D_{eff}$.

Figure 1 shows experimental bimFCS data for all three types of diffusion behavior. The $\tau_{1/2}$ versus $\omega^2$ relationship obtained from the fluorescently labeled lipid TAMRA-PE (carboxy-tetramethyl-rhodamine labeled phosphatidylethanolamine) diffusing in a supported POPC lipid bilayer is consistent with free Brownian motion in a homogenous medium with $D_{eff}$ = 2.3 ± 0.1 µm$^2$/s (at 25° C). Likewise, the diffusion of the fluorescent lipipophilic dye DiI (octadecyl-indocarbocyanine) in the cell membrane of HEK 293 indicates free Brownian motion with $D_{eff}$ = 1.33 ± 0.03 µm$^2$/s (at 25° C, n = 3) **(Fig. 1 c)**. In contrast, the $\tau_{1/2}$ versus $\omega^2$ relationship obtained from Glycosylphosphatidylinositol anchored GFP (GPI-GFP) in PtK2 cells has a positive time-axis intercept ($t_i$ = 17.4 ± 5.6 ms, T = 25° C) indicating that this protein interacts transiently with submicroscopic domains **(Fig. 1 d)**. Following cholesterol extraction from the cell membrane using methyl-β-cyclodextrin (β-CD), the time-axis intercept of the $\tau_{1/2}$ versus $\omega^2$ graph of GPI-GFP is reduced to zero ($t_i$ = 1.4 ± 6.9 ms, n = 6), now indicating free diffusion. This data is consistent with GPI-GFP being a 'lipid-raft associated protein' which is transiently trapped in cholesterol stabilized domains. The effective diffusion coefficient of GPI-GFP remains unchanged before and after cholesterol extraction, 0.62 ± 0.07 µm$^2$/s and 0.7 ± 0.1 µm$^2$/s, respectively (25° C). For the transferrin receptor (TfR) we observe a kink in the $\tau_{1/2}$ versus $\omega^2$ relationship and measure a negative time-axis intercept of $t_i$ = -30.0 ± 35.7 ms in PtK2 cells. The diffusion constant for TfR's long range diffusion is 0.22 ± 0.04 µm$^2$/s, while its short range diffusion is faster and similar to the GPI-GFP diffusion (25° C). This behavior is expected because TfR contains a large cytoplasmic domain, which has previously been shown to interact with 'membrane fences', i.e. membrane associated cytoskeletal filaments[2,15]. Therefore, our method also readily yields structural information about the density of the membrane cytoskeleton which is thought to be altered in a range of pathologies, i.e. cancer[16,17,20].

Most significantly, bimFCS is uniquely suited to characterize changes in a protein's association with lipid domains or the cytoskeleton in a single cell in real-time. This real-time capability allows investigation of the changes in receptor-lipid raft association caused by receptor activation and dimerization during cell signaling[3-5]. As a model of dimerization induced changes in protein-lipid raft association, we measured changes in GPI-GFP-lipid raft association in PtK2 cells following addition of a monoclonal anti-GFP antibody (**Figure 2 a, b**). **Figure 2 a** shows the $\tau_{1/2}$ versus $\omega^2$ relationship of GPI-GFP before and after dimerization by the antibody. Both curves exhibit a positive time axis intercept, consistent with lipid raft interaction. However, within the first minute of antibody induced dimerization the time axis intercept increased from 36 ± 4 ms to 51 ± 5 ms, indicating increased interaction of GPI-GFP with lipid rafts. The effective diffusion coefficient remained unchanged at 0.38 ± 0.02 µm$^2$/s before and 0.39 ± 0.02 µm$^2$/s after dimerization (PtK2 cells, 25° C). The time course and extent of increased lipid raft association is similar between individual cells (**Fig. 2 b**). Together, the 40% increase in the time-axis intercept with unchanged effective diffusion constant indicates the time spent in each domain increased while the total time spent in domains remained the same[18]. A likely explanation would be that several domains merged as consequence of the receptor dimerization, or alternatively that the rate of entering and of exiting the domains may have been decreased by a similar amount (see supplement).

Conformational changes in isolated proteins as consequence of activation may also modulate protein-lipid domain or protein-cytoskeleton interactions. Using bimFCS we quantified clustering of GFP labeled transient receptor vanilloid channel (TRPV2-GFP) during activation by 2-Aminoethoxydiphenyl borate (2-APB) in PtK2 cells (**Fig. 2 c**). The $\tau_{1/2}$ versus $\omega^2$ relationship for the resting, closed TRPV2-GFP displays a kink at $\omega^2$ = 0.13 $\mu m^2$, with the linear fit of the data of the smaller areas having a positive time-axis intercept. This indicates that diffusing, closed channels associate transiently with submicroscopic domains *and* collide with membrane fences. After TRPV2-GFP activation with the agonist 2-APB (50 $\mu M$), the $\tau_{1/2}$ versus $\omega^2$ relationship is shallower and no longer displays the kink, yet still has a positive time-axis intercept. However, the $\tau_{1/2}$ versus $\omega^2$ analysis only includes diffusing channels, while the FCS curves provide information on diffusing and immobile channels. **Fig. 2 d** shows that after 2-APB stimulation (right) all FCS curves obtained for the different areas $\omega^2$ decrease with two time-constants: one, which increases with increasing area $\omega^2$ and a second area-independent time-constant in the order of seconds. A decrease of all FCS curves independent of area size $\omega^2$ indicates a diffusion independent reduction of the fluorescence intensity, such as bleaching of an immobile protein fraction. Hence, the bimFCS data suggest that APB activation of GFP-TRPV2 causes channels which were previously 'hopping' across membrane fences, to anchor to the cytoskeleton. Several members of the TRP channel family are known to interact with the cytoskeleton and TRPV2 contains a large cytoplasmic domain, including an ankyrin repeat domain[21-22]. BimFCS provides now a way to study changes in this interaction during activation. We expect the bimFCS method to be especially useful in resolving how changes in membrane ultra-structure influence cell signaling.

## METHODS

Methods and any associated references are available in the online version of the paper (here they are attached at the end).

**Acknowledgments**
We thank J. Sankaran and T. Wohland for initial help with FCS analysis software development, T. Wohland, V. Rana and Y. Hsu for discussions, S.D. Parker and J. Pazik for molecular biology assistances, and S.D. Parker for editing.

**Author Contributions**
A.P. designed the study and supervised the project; H.H. conducted the experiments, wrote the software and analyzed the data; H.H. and A.P. wrote the manuscript.

# FIGURE LEGEND

**Figure 1 | Characterization of membrane protein diffusion by bimFCS.** (**a**) Schematic showing the diffusion of a membrane protein and how it may be reduced by either transient trapping in sub-microscopic domains (small green areas), or interaction with diffusion barriers, i.e. 'membrane fences' (red grid lines). The pixels of the EMCDD camera used to measure the diffusion are overlaid as square grid (grey). (**b**) Normalized FCS curves obtained with increasing pixel-bin size from a fluorescence image sequence of Rhodamine-labeled PE lipids diffusing in an artificial lipid bilayer. The shift of the curves along the $\tau$ axis indicates an increase in transit time $\tau_{1/2}$ with increasing observation area $\omega^2$. (**c**) $\tau_{1/2}$ versus $\omega^2$ relationships of Rhodamine-PE in a supported DPPC lipid bilayer and of the lipid dye DiI in the PtK2 cell membrane (**25° C**). Both curves intercept the time axis at the origin, indicating free Brownian motion. (**d**) Non-Brownian diffusion is observed for GFP-labeled membrane proteins in intact cells. The $\tau_{1/2}$ versus $\omega^2$ relationship of GPI-GFP (red Δ) displays a positive time-axis intercept. Upon cholesterol extraction using β-CD the time-axis intercept is reduced to zero (green ○). The $\tau_{1/2}$ versus $\omega^2$ curve for TfR-GFP (blue □) exhibits a kink around bin 2 and the trend line for the larger areas (bin 3-6) has a negative intercept, indicating a strong interaction with the membrane cytoskeleton 'fences'.

**Figure 2 | Dynamic changes in protein-domain association probed on single cells by bimFCS**. (**a**) The $\tau_{1/2}$ versus $\omega^2$ relationship of GPI-GFP before and after perfusion of αGFP antibody obtained in the same 8μm×3.2μm of a single PtK2 cell. The linear fit (dashed lines, bin 2 through bin 6) shows that the treatment caused an increase in time-axis intercept but left the effective diffusion $D_{eff}$ unchanged. (**b**) Kinetics of the change of the time-axis intercept in response to the addition of αGFP antibody (perfused on the cells at t=0 min for 8s) recorded in two different cells from two separate cultures. (**c**) The $\tau_{1/2}$ versus $\omega^2$ relationship of TRPV2-GFP obtained from the same cell before and after perfusion of agonist 2-APB (50 μM). **d**, bimFCS curves of TRPV2-GFP before and 2 min after 2-APB addition. The bimFCS curves obtained before stimulation (left) were fitted using a one-component Hill Equation (solid lines), while the data obtained after stimulation required a two-component Hill Equation fitting (right) because of the appearance of a photobleaching immobile population of molecules.

# FIGURE 1

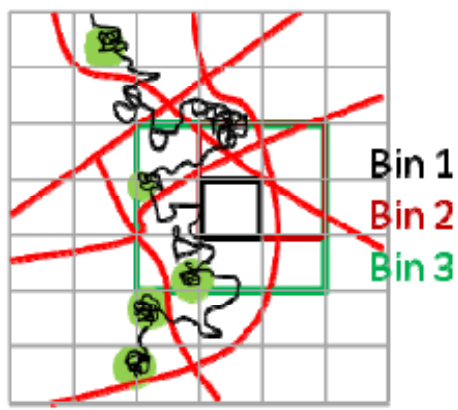
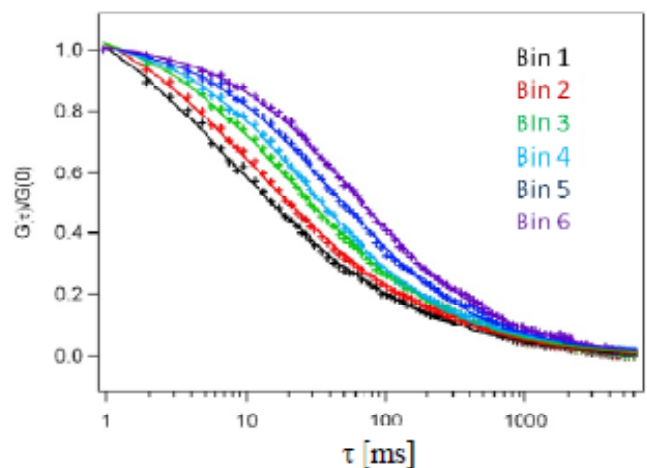
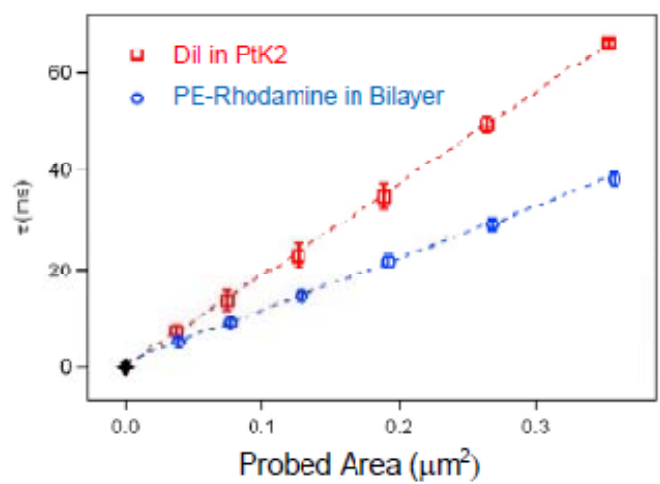
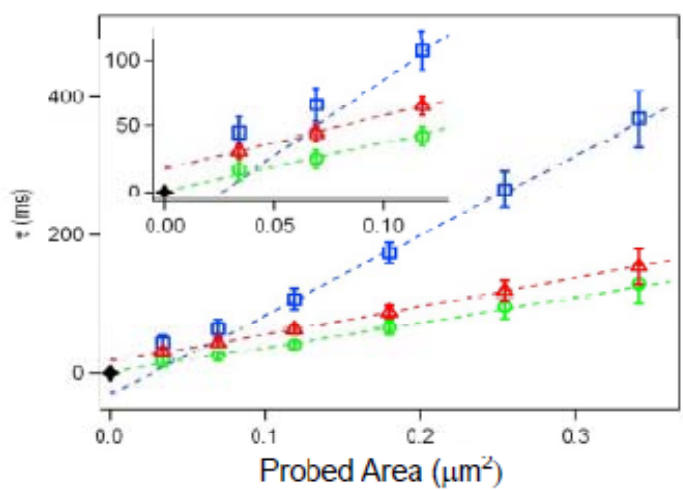

# FIGURE 2

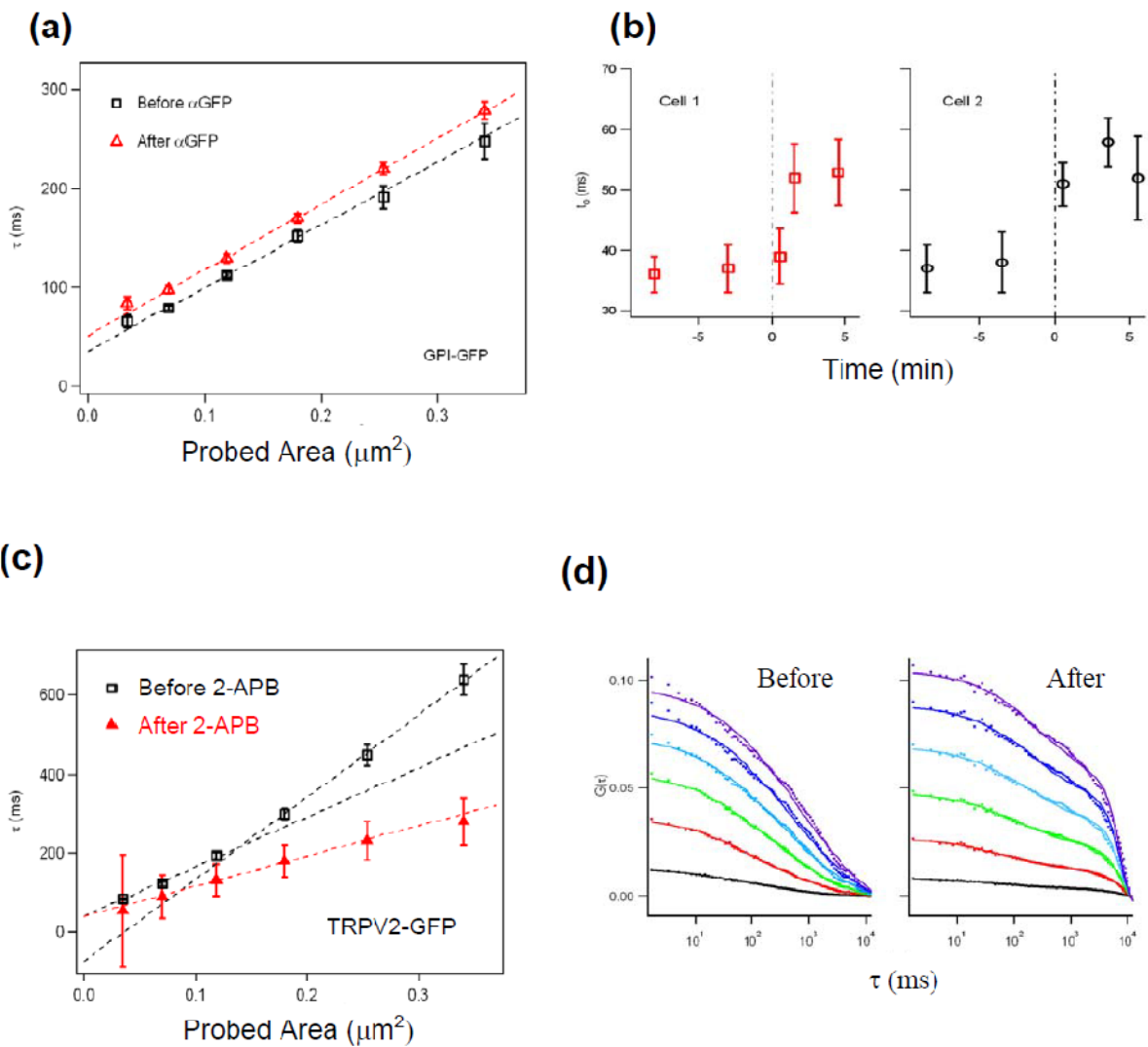

# REFERNCES

1. K. Jacobson, E. D. Sheets, and R. Simson, *Science* 268 (5216), 1441 (1995).
2. A. Kusumi, C. Nakada, K. Ritchie et al., *Annu Rev Biophys Biomol Struct* 34, 351 (2005).
3. Kai Simons and Winchil L.C. Vaz, *Annual Review of Biophysics and Biomolecular Structure* 33 (1), 269 (2004).
4. D. Lingwood and K. Simons, *Science* 327 (5961), 46 (2010).
5. Kai Simons and Elina Ikonen, *Nature* 387 (6633), 569 (1997).
6. R. Lasserre, X. J. Guo, F. Conchonaud et al., *Nat Chem Biol* 4 (9), 538 (2008).
7. J. F. Hancock, *Nat Rev Mol Cell Biol* 7 (6), 456 (2006).
8. A. Kusumi and K. Suzuki, *Biochim Biophys Acta* 1746 (3), 234 (2005).
9. K. Simons and D. Toomre, *Nat Rev Mol Cell Biol* 1 (1), 31 (2000).
10. Aiwei Tian, Corinne Johnson, Wendy Wang et al., *Physical Review Letters* 98 (20), 208102 (2007).
11. T. Harder, P. Scheiffele, P. Verkade et al., *J Cell Biol* 141 (4), 929 (1998).
12. A. Pralle, P. Keller, E. L. Florin et al., *J Cell Biol* 148 (5), 997 (2000).
13. P. Sharma, R. Varma, R. C. Sarasij et al., *Cell* 116 (4), 577 (2004).
14. R. Varma and S. Mayor, *Nature* 394 (6695), 798 (1998).
15. N. Morone, C. Nakada, Y. Umemura et al., *Methods Cell Biol* 88, 207 (2008).
16. C. Khanna, X. Wan, S. Bose et al., *Nat Med* 10 (2), 182 (2004).
17. J. T. Parsons, A. R. Horwitz, and M. A. Schwartz, *Nat Rev Mol Cell Biol* 11 (9), 633 (2010).
18. L. Wawrezinieck, H. Rigneault, D. Marguet et al., *Biophys J* 89 (6), 4029 (2005).
19. B. Kannan, L. Guo, T. Sudhaharan et al., *Anal Chem* 79 (12), 4463 (2007).
20. S. K. Patra, *Biochim Biophys Acta* 1785 (2), 182 (2008).
21. C. B. Phelps, R. R. Wang, S. S. Choo et al., *J Biol Chem* 285 (1), 731 (2010).
22. Goswami and T. Hucho, *FEBS J* **275** (19), 4684 (2008)

# SUPPLEMENTARY INFORMATION

## ONLINE METHODS

### Cell Culture

Human embryonic kidney (HEK 293) cells and PtK2 cells were cultured in Dulbecco's modified Eagle's medium (DMEM) supplemented with 10% fetal bovine serum and 1% penicillin-streptomycin (Invitrogen) at 37°C under 5% $CO_2$. For imaging, the cells were plated sparsely on 18-mm glass coverslips. Transfections were performed 24 h after plating using Lipofectamine 2000 (Invitrogen). Imaging was done within 24-36 h after transfection in serum-free, phenol-red-free DMEM supplemented with 10 mM HEPES.

### Supported Lipid Bilayer

1-palmitoyl-2-oleoyl-sn-glycero-3-phosphocholine (POPC), and 1,2-dioleoyl-sn-glycero-3-phosphoethanolamine-N-(lissamine rhodamine B sulfonyl) (Liss Rhod PE) were purchased from Avanti. Glass coverslips (Carolina Biological) were first treated with Piranha solution (3:1 concentrated Sulfuric acid to 30% hydrogen peroxide solution) at 65°C for 30 to 60 min then rinsed with 18MΩ water. They were stored in air. Two mixtures of lipids in chloroform ((1) POPC with 2.5% Liss Rhod PE and (2) only POPC) were dried with a stream of nitrogen and placed under vacuum for 1 hour. Lipids then were hydrated in 1x PBS (1 mg lipids / 1 ml PBS). The resulting vesicles were put through five freeze (liquid nitrogen) - thaw (37 °C water bath) cycles and then extruded a minimum of 20 times through polycarbonate membranes of 100 nm diameter pores (Avanti Mini-Extruder). Vesicle solution containing 2.5% dye-labeled lipids was mixed with vesicle solution of un-labeled lipids at 1:100 ratio and then sandwiched between a coverslip and a unifrost microscope slide for 1 to 2 min. Bilayers were formed by vesicle fusion on the glass coverslip surfaces.

### Data Acquisition

See Fig. S1 for a schematic diagram of the bimFCS instrument. All measurements were performed using an objective type TIRF microscope constructed around an inverted microscope (AxioObserver, Zeiss) and an Andor EMCCD camera working in kinetic acquisition mode. A 100x oil immersion objective (Zeiss, NA=1.45) was used to achieve TIRF angle illumination. Each camera pixel corresponds to a 160nm × 160nm are in the sample plane. Typically a rectangular-shaped region of interest (ROI) is chosen around 3μm from the edge of cell. Maximum frame rates would vary depending on the number of rows utilized for the ROI. A typical frame rate for a 50×20 pixels ROI positioned is around 1.7 ms/frame. For each data set, 70s of data was recorded in a selected ROI. To optimize the signal-to-noise ratio and photobleaching for the GFP-labeled proteins we used about 3mW laser power (measure at the objective lens, 488nm, Innova 70C laser (Coherent).

## Data Analysis

All data analysis was performed using Igor Pro software with self-written extension files. See Fig. S2 for a flow diagram of the data analysis. Fluorescence image stacks in the form of 8-bit tiff-files were loaded into a 3-D intensity matrix. The first 20s of each data set was discarded to avoid the initial uncertainties arising from the cascade effect in the EMCCD. Bleach correction was then performed in order to compensate the loss of fluorophores during data acquisition. For detailed protocol of bleach correction see "Bleach Correction". For each bin n (n = 1 – 6), the intensity values within each n × n pixel box is averaged, with the box sliding 1 pixel at a time until covering the whole frame. The FCS curve of each binned unit was calculated and then average over the frame. The averaged FCS function was mostly fitted with one component Hill Equation to obtain $\tau_{1/2,\text{bin N}}$. In cases where more than one dominant decay mechanism was observed, a global fitting of all 6 data sets with two-component Hill Equation was employed, with the longer component linked to be the same for all 6 bin settings. Effective observation area $\omega^2_{\text{binN}}$ was calculated taking into account the point spread function (PSF) of the system: $\omega^2_{\text{binN}} = [0.5*N*(\text{pixel size}) + 0.5* \omega_{\text{PSF}}]^2$, in which the $\omega_{\text{PSF}}$ is the width of the Gaussian function that approximates the system PSF obtained through calibration experiments. Monte-Carlo simulations of diffusion were performed to confirm the effective pixel size. The fluorescence of 100 simulated lipids diffusing for 120seconds was convoluted with the PSF and added together to create sample data with known molecule and intensity distribution.

## Interpretation of Changes in Time-axis Intercept and Effective Diffusion Coefficient

Wawrezinieck et al. derived that the time-axis intercept $t_0$ of a protein with a fraction $\alpha$ of molecules inside permeable micro-domains, can be approximated from the confinement time $\tau_{\text{conf}}$ in a single domain which is the time for a molecule in the center of the domain to escape, and from the free diffusion time $\tau_d^{\text{domain}}$ in an area equal to the domain (Wawrezinieck, 2005):

$$t_0 \sim 2\alpha \, ( \tau_{\text{conf}} - \tau_d^{\text{domain}} ) .$$

The effective diffusion coefficient in the presence of permeable microdomains is given by (Wawrezinieck, 2005):

$$D_{eff} = \begin{cases} (1-\alpha)D_{obst} & if \ \alpha < 0.5 \\ (1-\alpha)D_{out} & if \ \alpha > 0.5 \end{cases},$$

where $D_{\text{obst}}$ is the diffusion coefficient among impermeable obstacles and $D_{\text{out}}$ is the diffusion coefficient outside of the domains.

This means that if $t_0$ increases while $D_{\text{eff}}$ remains unchanged, then likely the domain size increases, but fraction of molecules in and outside of domains remain the same. Contrary, if $t_0$ increases and $D_{\text{eff}}$ decreases, then more molecules partition into the domains or spend more time inside the domains.
.

**Figure S1** Instrumental Setup

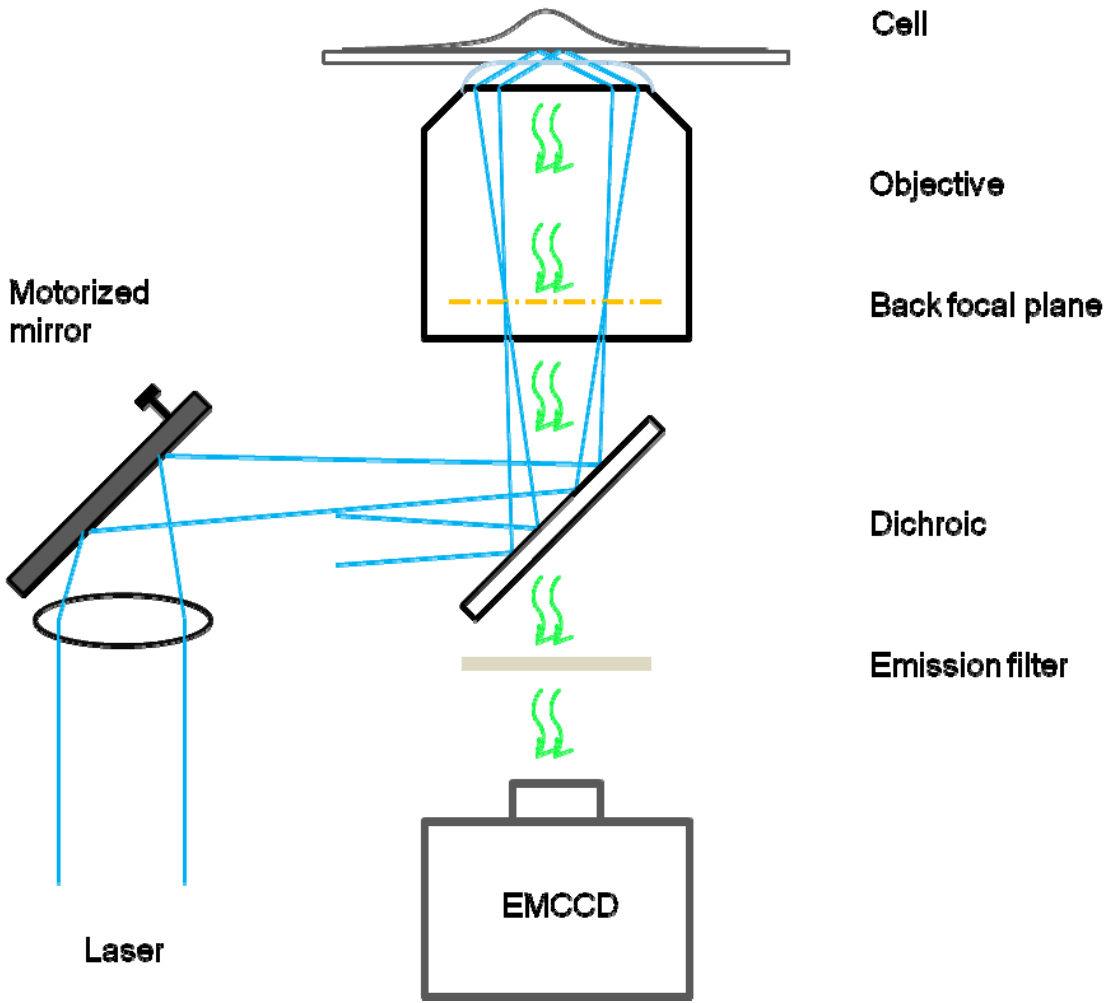

The 488nm line of an Argon-Krypton laser is focused on the back focal plane of the objective lens and steered by a motorized mirror in order to achieve TIRF angle. Fluorescence signals from the bottom membrane of the cell (or lipid bilayer) are collected by the 100X oil immersion objective, filtered and imaged by the EMCCD. Acquisition rates used are typically 0.5~2 ms/frame and the total acquisition time is 70s.

**Figure S2** Software Flow Diagram

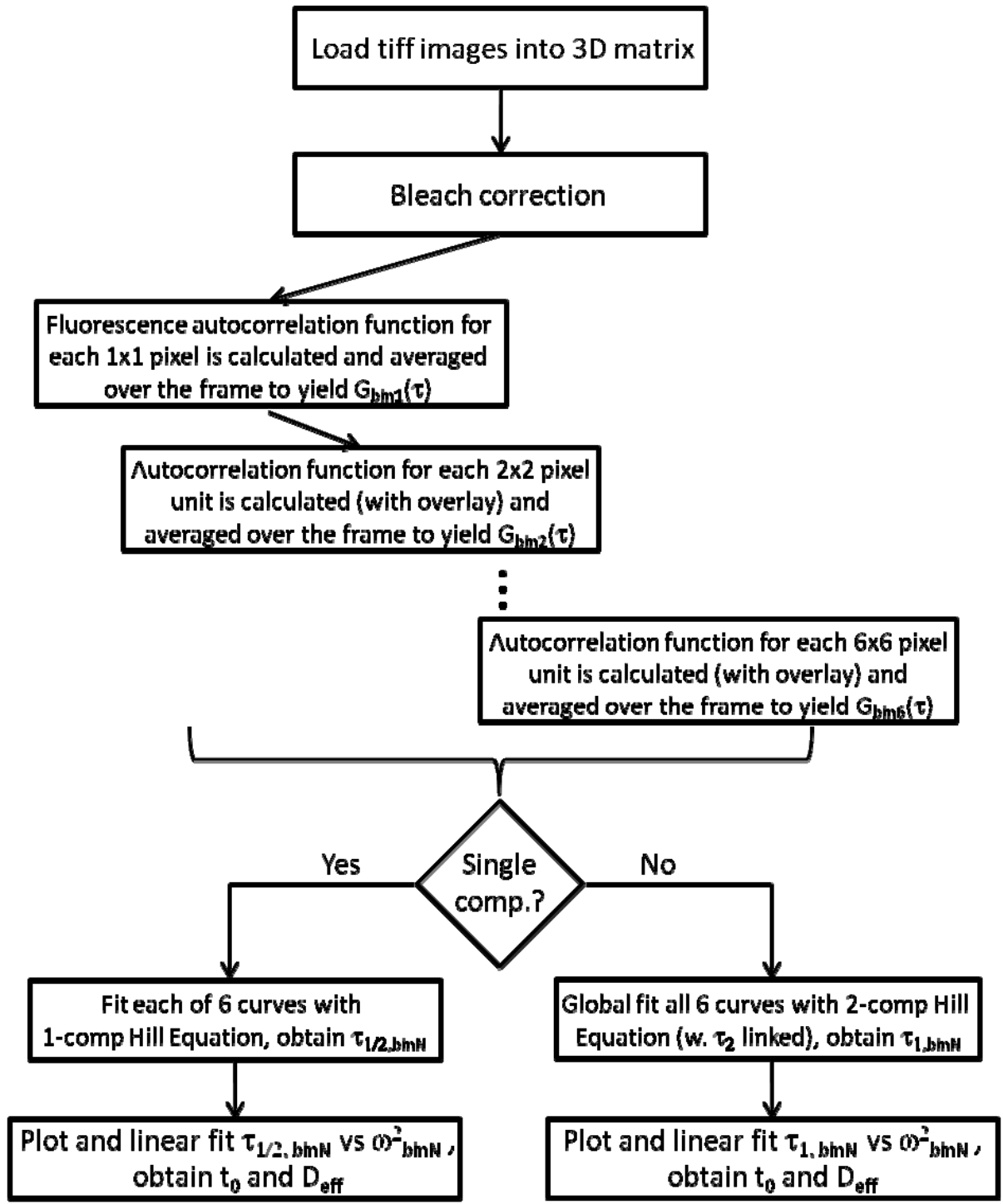

## Figure S3 Bleach correction

Once the image stack was loaded into a 3-D matrix, the average intensity of each frame ($I_{ave}$) is calculated and the resultant $I_{ave}(t)$ was fitted with an exponential decay function. The intensity value of each individual pixel is then divided by the average intensity value (when the ROI covers more than 100 pixels. Otherwise the fitted function of the $I_{ave}(t)$ was used for bleach correction) of the corresponding frame it lies in to correct for the effects of photobleaching, laser fluctuations, etc.

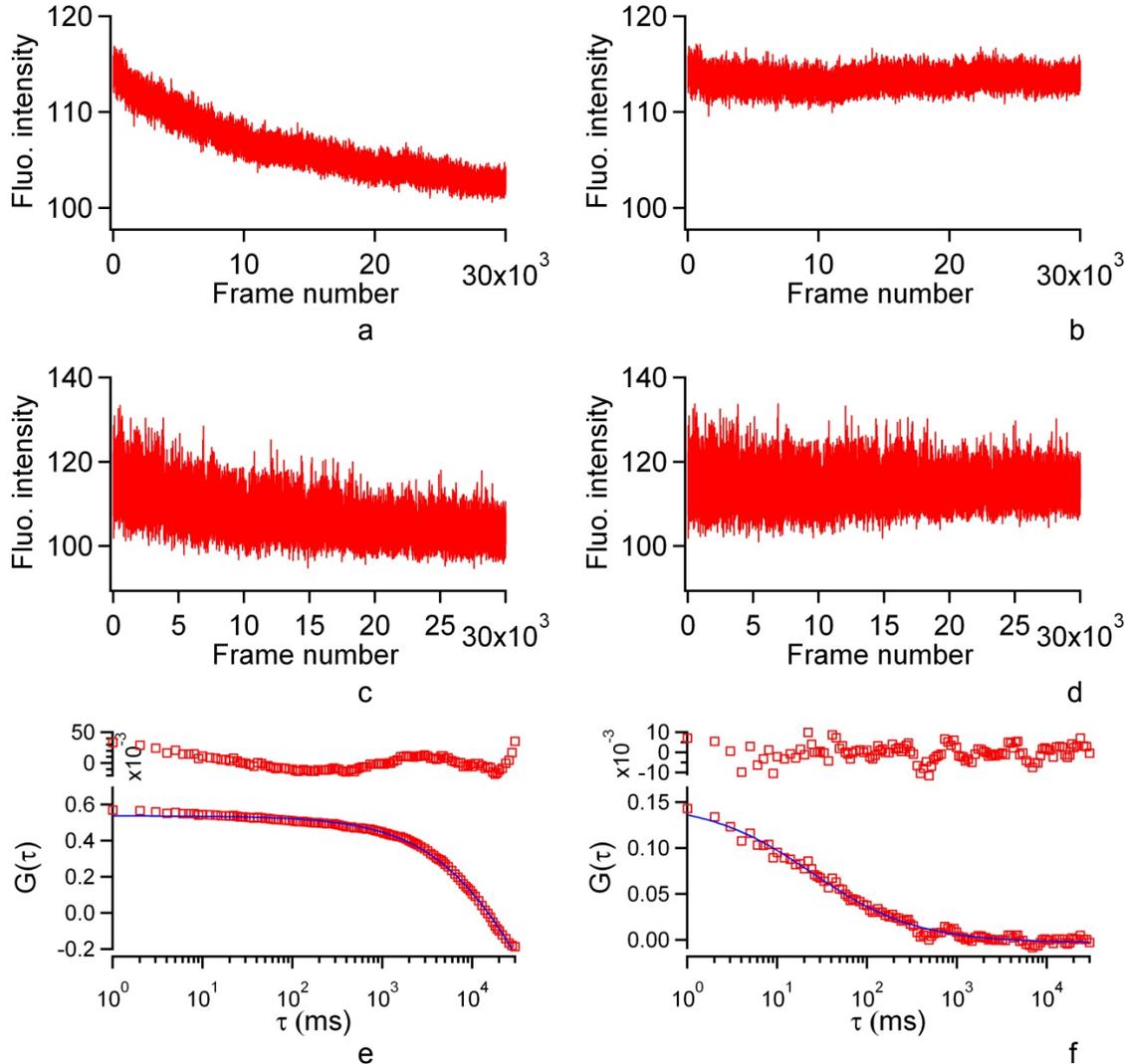

An example of the effects of bleach correction is given in Fig. S3. Data presented here was taken from a lipid bilayer in which 0.025% of the PE lipids were labeled with Rhodamine. The average frame intensities $I_{ave}$ before (a) and after bleach correction (b) were plotted over number of frames. Fluorescence fluctuations in a 3×3 binned area before and after bleach correction are shown c and d, respectively. A comparison of the resultant FCS curves shows that: without bleach correction, photobleaching (in the timescale of seconds) was the dominant decay mechanism (e); the shorter decay time due to diffusion (in tens of ms) only becomes apparent after bleach correction was performed (f). One component Hill Equation fit (blue lines) gives a decay time of $\tau_{1/2} = 50 \pm 10$ s and $\tau_{1/2} = 25 \pm 2$ ms for the FCS curves in e and f, respectively.